\documentclass{ws-ijmpd}
\usepackage{epsf}

\newcommand{\beq}{\begin{equation}}    % for lazy typers
\newcommand{\eeq}{\end{equation}}
\newcommand{\beqa}{\begin{eqnarray}}
\newcommand{\eeqa}{\end{eqnarray}}

\catcode`@=11     % make @ like a letter

\def\eqalign#1{\null\,\vcenter{\openup\jot\m@th
  \ialign{\strut\hfil$\displaystyle{##}$&$\displaystyle{{}##}$\hfil
      \crcr#1\crcr}}\,}
\def\meqalign#1{\null\,\vcenter{\openup\jot\m@th
  \ialign{\strut\hfil$\displaystyle{##}$&&$\displaystyle{{}##}$\hfil
      \crcr#1\crcr}}\,}

\def\imeqalign#1{\null\,\vcenter{\openup\jot\m@th
  \ialign{\strut\hfil$\displaystyle{##}$&$\displaystyle{{}##}$\hfil
         &&\hfil$\displaystyle{##}$&$\displaystyle{{}##}$\hfil
      \crcr#1\crcr}}\,}

\def\cmeqalign#1{\null\,\vcenter{\openup\jot\m@th
  \ialign{\strut\hfil$\displaystyle{##}$&&\hfil$\displaystyle{{}##}$\hfil
      \crcr#1\crcr}}\,}

\catcode`@=12   % back to nonletter category
\def\dual{\,{}^\ast{}\kern-1.5pt}
\def\dualu{\,{}^{\ast (u)}{}\kern-1.5pt}
\def\leftdual{\,{}^\ast{}\kern-1.5pt}

\def\dualp#1{{}^{\ast_{(\hbox{$\scriptstyle #1$})}} \kern-1pt}

\def\four{{}^{(4)}\kern-1pt}
\newcommand{\Tr}{\rm{Tr}}
\def\oversymbol#1#2{\vbox{\ialign{##\crcr \hfil$#1$\hfil\crcr
   \noalign{\kern1pt\nointerlineskip}%
   \hbox{$\hfil#2\hfil$}\crcr}}}
\def\overcirc#1{\oversymbol{\scriptstyle\kern.5pt \circ}{#1}}

\newcommand{\apj}{{\it Astroph. Jour.}}
\newcommand{\prl}{{\it Phys. Rev. Lett.} }
\newcommand{\prd}{{\it Phys. Rev.} {\bf D}}
\begin{document}

%%%%%%%%%%%%%%%%%%%%%%%%%%%%%%%%%%%%%%%%%%%%%%%%%%
\def\nocropmarks{\vskip5pt\phantom{cropmarks}}
\let\trimmarks\nocropmarks
%%% Pls. remove the comment sign (%) to switch off the trimmarks

%\markboth{C. Cherubini, D. Bini, S. Capozziello, R. Ruffini}
%{Second Order Scalar Invariants of the Riemann Tensor: Applications to Black Hole Spacetimes}

%%%%%%%%%%%%%%%% Publisher's Area please ignore %%%%%%%%%%%%%%%
%
\catchline{}{}{}
%
%%%%%%%%%%%%%%%%%%%%%%%%%%%%%%%%%%%%%%%%%%%%%%%%%
\title{SECOND ORDER SCALAR INVARIANTS \\
OF THE RIEMANN TENSOR:\\
APPLICATIONS TO BLACK HOLE SPACETIMES}

\author{CHRISTIAN CHERUBINI\footnote{cherubini@icra.it}}

\address{Institute of Cosmology and Gravitation, Portsmouth University, Portsmouth, PO12EG, UK and\\
Physics Department ``E.R.
Caianiello,'' University of Salerno, I--84081, Italy and \\
International Center for Relativistic Astrophysics,
I.C.R.A., \\
University of Rome La Sapienza,
I--00185 Roma, Italy
}

\author{
DONATO BINI\footnote{binid@icra.it}}

\address{
Istituto per Applicazioni della Matematica C.N.R., Naples, I--80131, Italy and\\
International Center for Relativistic Astrophysics,
I.C.R.A., \\
University of Rome La Sapienza,
I--00185 Roma, Italy}

\author{SALVATORE CAPOZZIELLO\footnote{capozziello@sa.infn.it}}

\address{Physics Department ``E.R. Caianiello,''
University of Salerno, I--84081, Italy
}

\author{REMO RUFFINI\footnote{ruffini@icra.it}}

\address{
Physics Department, University
of Rome La Sapienza,
I--00185 Roma, Italy\\
International Center for Relativistic Astrophysics,
I.C.R.A., \\
University of Rome La Sapienza,
I--00185 Roma, Italy}

\maketitle


\begin{abstract}
We discuss the Kretschmann, Chern-Pontryagin and Euler invariants among the  second order scalar invariants of the Riemann tensor in any spacetime in the Newman-Penrose formalism and in the framework of gravitoelectromagnetism, using the Kerr-Newman geometry as an example.
An analogy with electromagnetic invariants leads to the definition of regions of gravitoelectric or gravitomagnetic dominance.
\end{abstract}

\section{Introduction}

This article shows how the most interesting second order scalar invariants of the Riemann tensor (i.e.
the Kretschmann, Chern-Pontryagin and Euler invariants) among the set of 14 independent curvature scalar invariants can be expressed simply in any spacetime~\cite{gendeb,Witten,Petrov,rinpen,carmi}
not only in the Newman-Penrose formalism (NP), as is well known in the literature, but also in the framework of gravitoelectromagnetism (GEM), the $1+3$ splitting of spacetime. On the other hand,
working instead with standard tensor algebra, the evaluation of such invariants involves long calculations
for which a computer algebra system is often necessary.\cite{lake}
This widely held opinion is also reflected in a recent paper by Henry,\cite{Henry} which led to the present discussion and motivated further study.

However, within the NP  formalism, evaluating the second order scalar invariants of the Riemann tensor
is not so difficult. Furthermore in special cases (Petrov type D spacetimes for example), this is almost trivial
because the symmetries of the spacetime  can be adapted to the NP frame (i.e. the two repeated principal null directions can be aligned with the two real null vectors of the NP tetrad), with the subsequent vanishing of most of the Riemann tensor components.

Similarly, adopting a GEM approach through a relative observer splitting of spacetime, the task of evaluating second order scalar invariants of the Riemann tensor can also be accomplished in a simple and elegant way which also allows their interpretation in complete analogy with the well known second order scalar invariants of the Maxwell $2$-form. This is an an original contribution of the present article leading to a new interpretation  of the regions in which such ``quadratic curvatures" become negative.

The following second order scalar invariants of the Riemann tensor will be studied:
the Kretschmann invariant ($K_1$),
and its ``dual" counterparts which were omitted in the discussion of Henry:\cite{Henry} the Chern-Pontryagin ($K_2$) and the Euler invariant ($K_3$).\cite{Obukhov-Hehl}
These are discussed in detail for a Kerr-Newman spacetime, which
is presently of interest in view of its possible astrophysical applications.\cite{Punsly}

Very recently such scalars have been considered in the context of  black hole collisions and perturbations by Baker and Campanelli,\cite{bakcam} who introduced a `speciality index' (involving second and third order invariants) which provides a measure of how distorted a black hole perturbation is from the background spacetime.
During recent decades curvature invariants have also been considered in the context of quantum gravity and for effective theories of gravity applied to cosmology.
In such cases, they play an important role:
a) for the one-loop level renormalization of gravity\cite{birrel} and
b) for producing inflationary behaviour in early universe
cosmology.\cite{starobinski}
Finally  they have also been studied in
the context of conformal gravity, where in the post-Newtonian approximation limit, they generate corrections to the standard Newtonian potential which might have astrophysical implications,\cite{mannheim1,mannheim2} as a consequence of certain new gravitational field equations in which the metric is coupled to matter fields via the Bach tensor.\cite{tsaheh}

\section{
Kretschmann, Chern-Pontryagin and Euler Invariants}

We will consider the following second order scalar invariants of the Riemann tensor:
\begin{eqnarray}
K_1 &=& R_{\alpha\beta\gamma\delta}R^{\alpha\beta\gamma\delta}, \nonumber \\
K_2 &=& [{}^*R{}]_{\alpha\beta\gamma\delta}R^{\alpha\beta\gamma\delta}, \nonumber \\
K_3 &=& [{}^*R{}^*]{}_{\alpha\beta\gamma\delta}R^{\alpha\beta\gamma\delta}.
\end{eqnarray}

Starting from the relation between the Riemann, Weyl and Ricci tensors (or first Matt\'e-decomposition of the Weyl
tensor:\cite{roberts})
\begin{eqnarray}
R_{\alpha\beta\gamma\delta}&=&C_{\alpha\beta\gamma\delta}+\frac{1}{2} ( g_{\alpha \gamma} R_{\beta \delta}- g_{\beta \gamma} R_{\alpha \delta}- g_{\alpha \delta}R_{\beta \gamma}+ g_{\beta \delta}R_{\alpha \gamma})\nonumber \\
&&-\frac{1}{6}( g_{\alpha \gamma}g_{\beta \delta}-g_{\alpha \delta}g_{\beta \gamma})R \ ,
\end{eqnarray}
one can write $K_1$, $K_2$ and $K_3$ in the form:
\begin{eqnarray}
\label{KAPPA1}
K_1&=&C_{\alpha\beta\gamma\delta}C^{\alpha\beta\gamma\delta}+2R_{\alpha\beta}R^{\alpha\beta}-\frac{1}{3}R^2\, \nonumber  \\
&=&I_1 + 2R_{\alpha\beta}R^{\alpha\beta}-\frac{1}{3}R^2\, ,\\
\label{KAPPA2}
K_2&=&[{}^*C{}]_{\alpha\beta\gamma\delta}C^{\alpha\beta\gamma\delta}\,  \nonumber \\
&=& I_2\, , \\
\label{KAPPA3}
K_3&=&-C_{\alpha\beta\gamma\delta}C^{\alpha\beta\gamma\delta}+2R_{\alpha\beta}R^{\alpha\beta}-\frac{2}{3}R^2\, \nonumber \\
   &=&-I_1+2R_{\alpha\beta}R^{\alpha\beta}-\frac{2}{3}R^2 .
\end{eqnarray}
where $I_1$ and $I_2$ are the only independent second order scalar invariants of the Weyl tensor:
\begin{eqnarray}
\label{I1}
I_1&=&C_{\alpha\beta\gamma\delta}C^{\alpha\beta\gamma\delta}\, , \\
\label{I2}
I_2&=&[{}^*C{}]_{\alpha\beta\gamma\delta}C^{\alpha\beta\gamma\delta}\, ,
\end{eqnarray}
because of the self-duality property of the Weyl tensor: $[{}^*C{}^*]=-C$. The relation $K_2=I_2$
can easily be derived
by using the tracefree property of the Weyl tensor and the Bianchi identities of the first kind for the Riemann tensor.
It is worthwhile noting the following properties of $K_1$, $K_2$ and $K_3$ and $I_1$ and $I_2$:
\begin{enumerate}
\item The sum
\begin{equation}
K_1+K_3=4R_{\alpha\beta}R^{\alpha\beta}-R^2 = 4\kappa^2 [T_{\alpha\beta}T^{\alpha\beta}-\frac14 T^2]\, ,
\end{equation}
as a consequence of the Einstein equations, cleanly separates off the non-Weyl part of the curvature determined by the matter content.
Analogously one has:
\begin{equation}
K_1-K_3 =2I_1 +\frac13 R^2 \, .
\end{equation}

\item
The density associated with $K_3$, i.e.
${\cal K}_3=\sqrt{-g}K_3$,
is proportional to the topological Euler density:
\begin{equation}
{\cal E}=-\frac{1}{128 \pi^2} {\cal K}_3=\frac{1}{128 \pi^2} \sqrt{-g}(R_{\alpha\beta\gamma\delta}R^{\alpha\beta\gamma\delta}-4R_{\alpha\beta}R^{\alpha\beta}+R^2)\, .
\label{EULDENS}
\end{equation}
Furthermore ${\cal K}_3$ is the  divergence of a vector density:
\begin{equation}
{\cal K}_3=-\partial_\alpha {\cal D}^\alpha\, ,
\end{equation}
where in coordinate components one has
\begin{equation}
{\cal D}^\alpha = \sqrt{-g}\, \eta^{\alpha\beta\gamma\delta}\eta_{\rho\sigma}{}^{\mu\nu}
\Gamma^\rho{}_{\mu\beta} [\frac12 R^\sigma{}_{\nu\gamma\delta} +\frac13 \Gamma^\sigma{}_{\lambda\gamma} \Gamma^\lambda{}_{\nu\delta} ]\, .
\end{equation}

\item $K_2$ and $K_3$ are two topological invariants obtained (in four dimensions)
from the curvature two-forms. The Gauss-Bonnet theorem states that the integral of $K_3$ over a compact manifold without boundary is proportional to its Euler characteristic. The integral of $K_2$ instead is related to the so called ``instanton number" of the manifold.\cite{Obukhov-Hehl}

\item
The variational derivative of the density associated with $I_1$,
i.e. ${\cal I}_1=\sqrt{-g}I_1$,
is the  Bach tensor\cite{hansj}
\begin{equation}
B^{\alpha\beta}=\frac{1}{\sqrt{-g}}\frac{\delta {\cal I}_1}{\delta g_{\alpha\beta}} \, ,\label{BACHEQNS}
\end{equation}
which is symmetric and tracefree.
The previous considerations concerning the topological invariants allow a simple and elegant way to calculate the Bach tensor.\cite{tsaheh}
In fact from Eq.~(\ref{KAPPA2}) it follows that
\begin{equation}
{\cal I}_1=\sqrt{-g}(R_{\alpha\beta\gamma\delta}R^{\alpha\beta\gamma\delta}-2R_{\alpha\beta}R^{\alpha\beta}+\frac13 R^2)\, ,
\end{equation}
which is equivalent to
\begin{eqnarray}
{\cal I}_1=128\pi^2 {\cal E}+\sqrt{-g}(2R_{\alpha\beta}R^{\alpha\beta}-\frac23 R^2)\,. \label{trick}
\end{eqnarray}
The first term in (\ref{trick}) is  the Euler density (\ref{EULDENS}), and consequently it does not contribute to Eq.~(\ref{BACHEQNS}).\cite{Parker} The variation of the second term gives directly the Bach tensor:
\begin{eqnarray}
   B_{\alpha\beta}&=&\frac{2}{3} \, R_{;\alpha;\beta} -  2\,{R_{\alpha\beta;\mu}}^{;\mu}
   +\frac{1}{3}\,g_{\alpha\beta}\,{R_{;\mu}}^{;\mu} -\frac{1}{3}\, R^2 g_{\alpha\beta}  \nonumber \\
   &&+ \frac{4}{3} \, R \, R_{\alpha\beta} + g_{\alpha\beta} \,
   R_{\mu\nu}\, R^{\mu\nu} - 4 \, R^{\mu\nu} \, R_{\alpha \mu\beta\nu}  .\label{BACHEQUATIONS}
\end{eqnarray}

\end{enumerate}

\subsection{Newman-Penrose Formalism}

The two second order scalar invariants of the Weyl tensor $I_1$ and $I_2$
satisfy the NP relation:\cite{sib}
\begin{equation}
I=I_1-iI_2=16(3\psi_2^2+\psi_0\psi_4-4\psi_1\psi_3),
\, \label{COLLO}
\end{equation}
where for the NP notation we follow the conventions of Chandrasekhar.\cite{chandra}
By taking the real and the imaginary parts (the latter reversed in sign) of Eq.~(\ref{COLLO}) one finds:
\begin{eqnarray}
&\,&I_1=\Re e(I)=8\left[(3\psi_2^2+\psi_0\psi_4-4\psi_1\psi_3)+\overline{(3\psi_2^2+
\psi_0\psi_4-4\psi_1\psi_3)}\right]\ \nonumber ,\\
&\,&I_2=-\Im m(I)=8i\left[(3\psi_2^2+\psi_0\psi_4-4\psi_1\psi_3)-\overline{(3\psi_2^2+
\psi_0\psi_4-4\psi_1\psi_3)}\right]\ .\label{WEYL1}
\end{eqnarray}
Projecting $R_{\alpha\beta}R^{\alpha\beta}$ onto the NP frame gives:
\begin{eqnarray}
R_{\alpha\beta}R^{\alpha\beta}&=&2[ \,{R_{11}}\,{R_{22}}+\,{{R_{12}}}^{2}-2\,{R_{13}}\,{R_{24}}-2\,
{R_{14}}\,{R_{23}}+\,{R_{33}}\,{R_{44}}+\,{{R_{34}}}^{2}]\nonumber\\
&=&8[\,\Phi_{{00}}\Phi_{{22}}+18\,{\Lambda}^{2}+2\,{\Phi_{{11}}}^{2}
-2\,\Phi_{{01}}\Phi_{{21}}-2\,\Phi_{{10}}\Phi_{{12}}+\,\Phi_{{0
2}}\Phi_{{20}}]\ .
\label{RICCY}
\end{eqnarray}
Finally $R$ in the NP formalism becomes:
\begin{equation}
R=24\Lambda .
\label{LAMBY}
\end{equation}
By inserting (\ref{WEYL1}), (\ref{RICCY}) and (\ref{LAMBY}) into (\ref{KAPPA1}), (\ref{KAPPA2}) and (\ref{KAPPA3}), we can express $K_1$, $K_2$ and $K_3$ in terms of the NP quantities in a form valid in any spacetime.

\subsection{Gravitoelectromagnetic Formalism}

A $1+3$ splitting of the spacetime is accomplished (locally) by introducing a family of test observers, i.e. a congruence of timelike lines with unit tangent vector $u$ ($u\cdot u=-1$).
$u$ defines both a local time direction and a local space (through  its orthogonal subspace in the tangent space).
Projecting tensor and  tensor equations along $u$ or orthogonally to $u$ defines  a `measurement process' associated with the observer family $u$; for example the `measurement' of a $1\choose1$ tensor $S$ results in a set of four spatial fields (i.e. for which any contraction by $u$ vanishes):
\begin{equation}
\{u^\delta u_\gamma S^\gamma{}_\delta, P(u)^\alpha{}_\gamma u^\delta S^\gamma{}_\delta,
P(u)^\delta{}_\alpha u_\gamma S^\gamma{}_\delta, P(u)^\alpha{}_\gamma P(u)^\delta {}_\beta S^\gamma{}_\delta \},
\end{equation}
where $P(u)^\alpha{}_\beta =\delta^\alpha{}_\beta+  u^\alpha u_\beta$ is the projector orthogonal to $u$.

Details about the splitting process systematically applied to spacetime tensor and spacetime differential operators as well as the origins of gravitoelectromagnetism can be found in Bini et al~\cite{mfg} to which we refer for notation and conventions.
Here we are interested in the splitting of the Riemann tensor.
The only nonvanishing spatial fields associated with the `measurement' of $R$ by the observer family $u$ are the following\cite{mtw}
\begin{eqnarray}
{\cal E}(u)_{\alpha\beta}&=& R_{\alpha\mu\beta\nu}u^\mu u^\nu , \nonumber \\
{\cal H}(u)_{\alpha\beta}&=& \frac12 \eta(u)^{\mu\nu}{}_\beta R_{\alpha\sigma \mu\nu}u^\sigma  , \nonumber \\
{\cal F}(u)_{\alpha\beta}&=& \frac14 \eta(u)^{\rho\sigma}{}_\alpha \eta(u)^{\mu\nu}{}_\beta R_{\rho\sigma\mu\nu} \ ;
\end{eqnarray}
${\cal E}(u)$, ${\cal H}(u)$ and ${\cal F}(u)$ have been respectively called the electric-electric, the electric-magnetic and the magnetic-magnetic parts of the Riemann tensor.
A straightforward calculation shows:
\begin{eqnarray}
\label{gravk}
K_1 &=& 4[\Tr {\cal E}(u)^2 -2 \Tr {\cal H}(u)\cdot {\cal H}(u)^T+ \Tr {\cal F}(u)^2]\, , \nonumber \\
K_2 &=& -8[\Tr {\cal H}(u)\cdot ({\cal E}(u)- {\cal F}(u) )]\, ,\nonumber \\
K_3 &=&  8 [\Tr {\cal E}(u)\cdot {\cal F}(u)+ \Tr {\cal H}(u)^2 ]\ .
\end{eqnarray}

\section{Application: The Kerr-Newman Spacetime}

The Kerr-Newman solution is of Petrov type D and in an NP frame adapted to the two repeated principal null directions (a Kinnersley tetrad~\cite{Kinnersley}) one has  only $\psi_2\neq 0$, so that eqs. (\ref{WEYL1}) become:
\begin{eqnarray}
&\,&I_1=24(\psi_2^2+ \overline{\psi_2^2})\,,\qquad
I_2=24i(\psi_2^2- \overline{\psi_2^2})\,.\label{WEYL000}
\end{eqnarray}
Moreover, as an electrovac solution of the Einstein-Maxwell system, the Kerr-Newman solution
has a tracefree Ricci tensor $R=0$ as a consequence of the tracefree property of the energy-momentum tensor of the electromagnetic source field.
Thus, in order to evaluate $K_1$, $K_2$ and $K_3$ we need only calculate $R_{\alpha\beta}R^{\alpha\beta}$
or the components of the Ricci tensor, namely $\Lambda$ and $\Phi_{mn}$.
However, here $\Lambda=0$ and  $\Phi_{mn}=2\phi_{m}\overline{\phi_{n}}$,  the only surviving  component of which is $\Phi_{11}$.
Thus
\begin{equation}
\Phi_{11}=-\frac{1}{4}(R_{12}+R_{34}),\qquad \Lambda=-\frac{1}{12}(R_{12}-R_{34})=0,
\end{equation}
which can be inverted yelding
\begin{eqnarray}
R_{12}=R_{34}=-2(\Phi_{11})=-4\phi_1\overline{\phi_1} .
\end{eqnarray}
By using these relations in eq. (\ref{RICCY}), we have:
\begin{equation}
R^{\alpha\beta}R_{\alpha\beta}=64 \phi_1^2 \overline{\phi_1}^2
\end{equation}
so that
\begin{eqnarray}
K_1=24(\psi_2^2+ \overline{\psi_2^2})+128(\phi_1\overline{\phi_1})^2\,.
\end{eqnarray}
where:\cite{Bose}
\begin{eqnarray}
&\,&\rho =-{(r-ia\cos\theta)}^{-1},\qquad \psi_2=\rho^3 (M+Q^2\,\overline{\rho})\,,\qquad \phi_1=\frac{1}{2}Q\rho^2\,.
\end{eqnarray}
The final result is:
\begin{equation}
R^{\alpha\beta}R_{\alpha\beta}=4\frac{Q^4}{(r^2+a^2\cos^2\theta)^4}
\end{equation}
and
\begin{eqnarray}
K_1&=&\frac{8}{\left (r^2+{a}^{2}\cos^2\theta\right )^{6}}\nonumber\\
&\times&[6{M}^{2}\left(\,{r}^{6}-15\,{r}^{
4}{a}^{2}\cos^2\theta+15\,{r}^{2}{a}^{4}\cos^4
\theta-\,{a}^{6}\cos^6\theta\right )\nonumber\\
&-&12M{Q}^{2}r\left({r}^{4}-10\,{r}^{2}{a}^{2}\cos^2\theta+5{a}^{4}
\cos^4\theta\right )\nonumber\\
&+&{Q}^{4}\left (7\,{r}^{4}-34\,{r}^{2}{a}^{2}
\cos^2\theta+7\,{a}^{4}\cos^4\theta\right )],
\end{eqnarray}
which coincides with the expression obtained by Henry\cite{Henry} using a  computer algebra system.
In the same way we can calculate $K_2=24 i (\psi_2^2- \overline{\psi_2^2})$:
\begin{eqnarray}
K_2&=&\frac {96a\cos\theta}{\left ({r}^{2}+{a}^{2}\cos^2\theta\right )^{6}}\nonumber\\
&\times&[{M}^{2}{r}\left (3{r}^{4}-10\,{r}^{2}{a}^{2}\cos^2\theta+3{a}^{4}\cos^4\theta\right )\nonumber\\
&-&M{Q}^{2}\left (5\,{r}^{4}-10{a}^{2}{r}^{2}\cos^2\theta+{a}^{4}\cos^4\theta\right )\nonumber\\
&+&2r{Q}^{4}\left({r}^{2}-a^2\cos^2\theta\right)]\, ,
\end{eqnarray}
and $K_3=-24(\psi_2^2+ \overline{\psi_2^2})+128(\phi_1\overline{\phi_1})^2$:
\begin{eqnarray}
K_3&=&-\,\frac{8}{\left (r^2+{a}^{2}\cos^2\theta\right )^{6}}\nonumber\\
&\times&[6{M}^{2}\left(\,{r}^{6}-15\,{r}^{
4}{a}^{2}\cos^2\theta+15\,{r}^{2}{a}^{4}\cos^4
\theta-\,{a}^{6}\cos^6\theta\right )\nonumber\\
&-&12M{Q}^{2}r\left({r}^{4}-10\,{r}^{2}{a}^{2}\cos^2\theta+5{a}^{4}
\cos^4\theta\right )\nonumber\\
&+&{Q}^{4}\left (5\,{r}^{4}-38\,{r}^{2}{a}^{2}
\cos^2\theta+5\,{a}^{4}\cos^4\theta\right )]\,.
\label{K3}
\end{eqnarray}
The expression of $I_1$ ($I_2\equiv K_2$) is instead
\footnote{According to the notation of the present paper, the quantity $K$ introduced by de Felice \cite{fdf}  as the ''first Khretshmann invariant"  actually is $I_1$.}
\begin{equation}
I_1=\frac{48}{(r^2+a^2\cos^2\theta)^6}(r^4+a^4\cos^4\theta-6a^2\cos^2\theta)(Q^2-Q_+^2)(Q^2-Q_-^2)
\end{equation}
with
\begin{equation}
Q^2_\pm=\frac{M(r\mp a\cos\theta)(r^2+a^2\cos^2\theta\pm 4ar\cos\theta)}{r^2-a^2\cos^2\theta \pm 2ar\cos\theta}\,.
\end{equation}

At large distances $K_1$, $K_2$ and $K_3$ have the expansion:\cite{Ciufolini}
\begin{eqnarray*}
&K_1&=48\,{\frac {{M}^{2}}{{r}^{6}}}-96\,{\frac {M{Q}^{2}}{{r}^{7}}}+56\,{\frac {{Q}^{4}-18\,{M}^{2}{a}^{2}\cos^2\theta
}{{r}^{8}}}+O\left(\frac{1}{r^9}\right)\, ,\nonumber\\
&\,&\nonumber\\
&K_2&=288\,{\frac {{M}^{2}a \cos\theta}{{r}^{7}}}-480\,{\frac {Ma\cos\theta{Q}^{2}}{{r}^{8}}}
+192\,
{\frac {a\cos\theta\left ({Q}^{4}-14\,{M}^{2}{a}^{2}\cos^2\theta\right )}{{r}^{9}}}+O\left(\frac{1}{r^{10}}\right)\, ,\qquad\qquad\nonumber \\
&K_3&=-48\,{\frac {{M}^{2}}{{r}^{6}}}+96\,{\frac {M{Q}^{2}}{{r}^{7}}}-8\,{\frac {5{Q}^{4}-126\,{M}^{2}{a}^{2}\cos^2\theta
}{{r}^{8}}}+O\left(\frac{1}{r^9}\right) .
\end{eqnarray*}
Instead, expanding for small $a/M$ and $Q/M$  one finds
\begin{eqnarray*}
&\,&K_1=48\,{\frac {{M}^{2}}{{r}^{6}}}-1008\,{\frac {{M}^{2}{a}^{2}\cos^{2}
\theta}{{r}^{8}}}-96\,{\frac {M{Q}^{2}}{{r}^{7}}}+...\simeq -K_3\nonumber\\
&\,&K_2=288\,{\frac {{M}^{2}a\cos\theta}{{r}^{7}}}-2688\,{\frac {{M}^{2}{a}^
{3}\cos^{3}\theta}{{r}^{9}}}-480\,{\frac {Ma\cos
\theta{Q}^{2}}{{r}^{8}}}+...
\end{eqnarray*}

It is also possible to find the surfaces on which $K_1$, $K_2$ and $K_3$ vanish.
Solving $K_2=0$ for $\cos \theta$ in the Kerr-Newman spacetime gives the following curves:
\begin{eqnarray}
&\,& \cos \theta=0 \, ,\nonumber \\
&\,& \cos \theta=\pm \frac{\sqrt{Mr(3Mr-2Q^2)}}{Ma} \, ,\nonumber \\
&\,& \cos \theta=\pm \frac{r\sqrt{(Mr-Q^2)(3Mr-Q^2)}}{a(3Mr-Q^2)}\, ,
\label{KNZEROK2}
\end{eqnarray}
which, in the Kerr case,  reduce to:
\begin{eqnarray}
&\,& \cos \theta=0 ,\nonumber \\
&\,& \cos \theta=\pm \frac{\sqrt{3}}{a} r,\nonumber \\
&\,& \cos \theta=\pm \frac{1}{\sqrt{3}a}r, \label{KERRZEROK2}
\end{eqnarray}
To solve $K_1=0$ it is convenient to re-express it in terms of the new variable $Z=a^2\cos^2 \theta$, yelding the cubic equation:
\begin{eqnarray}
&&\frac{8}{({r}^{2}+Z)^{6}}[
-6\,{M}^{2}{Z}^{3}
+(90\,{r}^{2}{M}^{2}-60\,Mr{Q}^{2}+
7\,{Q}^{4}){Z}^{2}\nonumber \\
&&\qquad +(-90\,{r}^{4}{M}^{2}+120\,M{r}^{3}{Q}^{2}
-34\,{r}^{2}{Q}^{4})Z+6\,{M}^{2}{r}^{6}
\nonumber \\
&&\qquad -12\,M{r}^{5}{Q}^{2}+
7\,{Q}^{4}{r}^{4}]=0\ ,
\end{eqnarray}
which can be solved exactly by using the Cardano formulas. Instead of giving these lengthy formulas here, we  give the simpler solutions for the Kerr case:
\begin{eqnarray}
\label{kerrzeros}
r&=&\pm a\cos\theta ,\nonumber \\
r&=&(2\pm\sqrt{3}) a\cos\theta , \nonumber \\
r&=&(-2\pm\sqrt{3}) a\cos\theta \ .
\end{eqnarray}
The equation $K_3=0$ may be handled similarly; for the simpler Kerr case it reduces to $K_3=-K_1=0$ and has the
same solutions.

\begin{figure}
\begin{center}
\epsfysize=2.8in \epsfbox{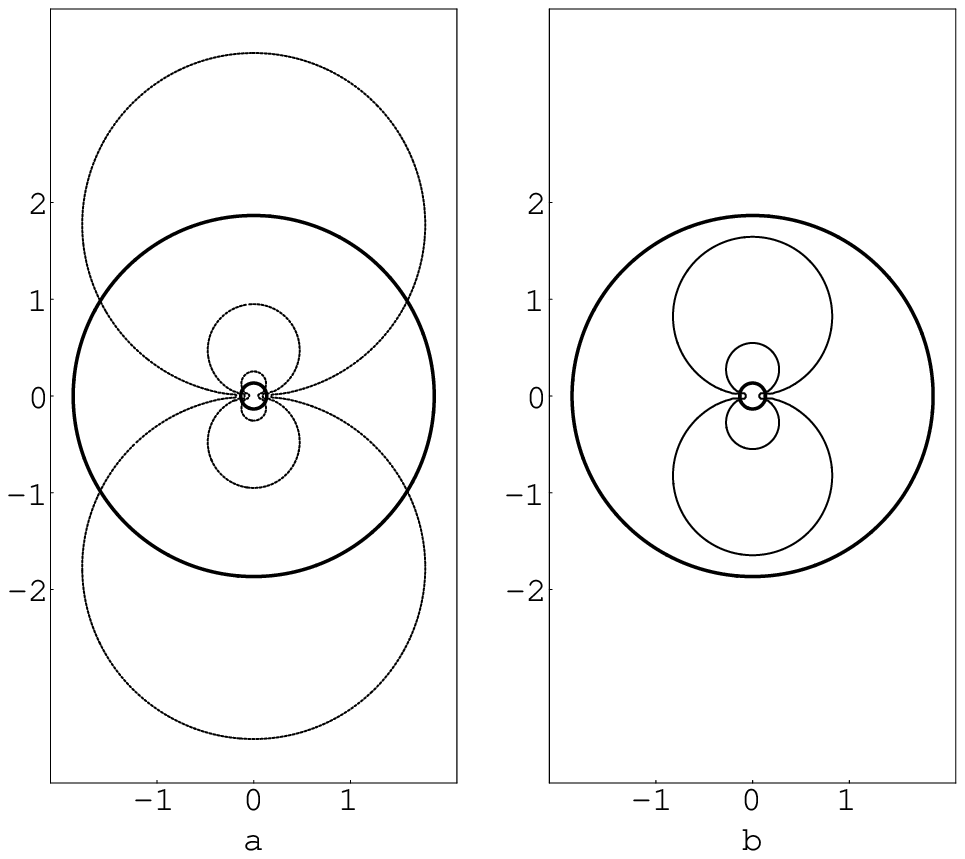}
\end{center}
\caption{
Kerr Black Hole with parameters $M=1$, $a=0.5$. Dashed lines in Fig. 1a correspond to both $K_1=0$ and $K_3=0$ ($K_1=-K_3$ in vacuum); thin solid lines in Fig. 1b correspond to $K_2=0$; thick solid lines in both Fig. 1a and Fig. 1b are the inner/outer horizons.
}
\label{fig1}
\end{figure}

\begin{figure}
\begin{center}
\epsfysize=5.0in \epsfbox{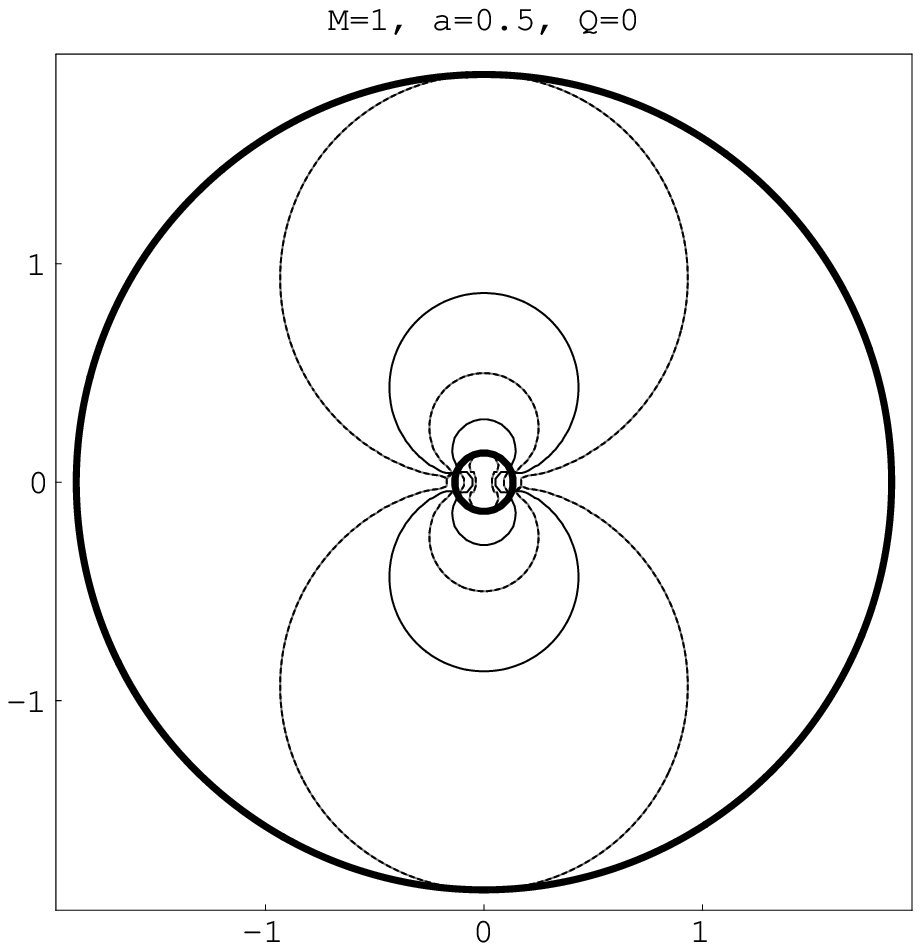}
\end{center}
\caption{
Kerr Black Hole with parameters $M=1$, $a=0.5$. Superposition of Figs. 1a and 1b. Dashed lines correspond to both $K_1=0$ and $K_3=0$ ($K_1=-K_3$ in vacuum); thin solid lines correspond to $K_2=0$; thick solid lines are the inner/outer horizons.
}
\label{fig2}
\end{figure}

\begin{figure}
\begin{center}
\epsfysize=2.8in \epsfbox{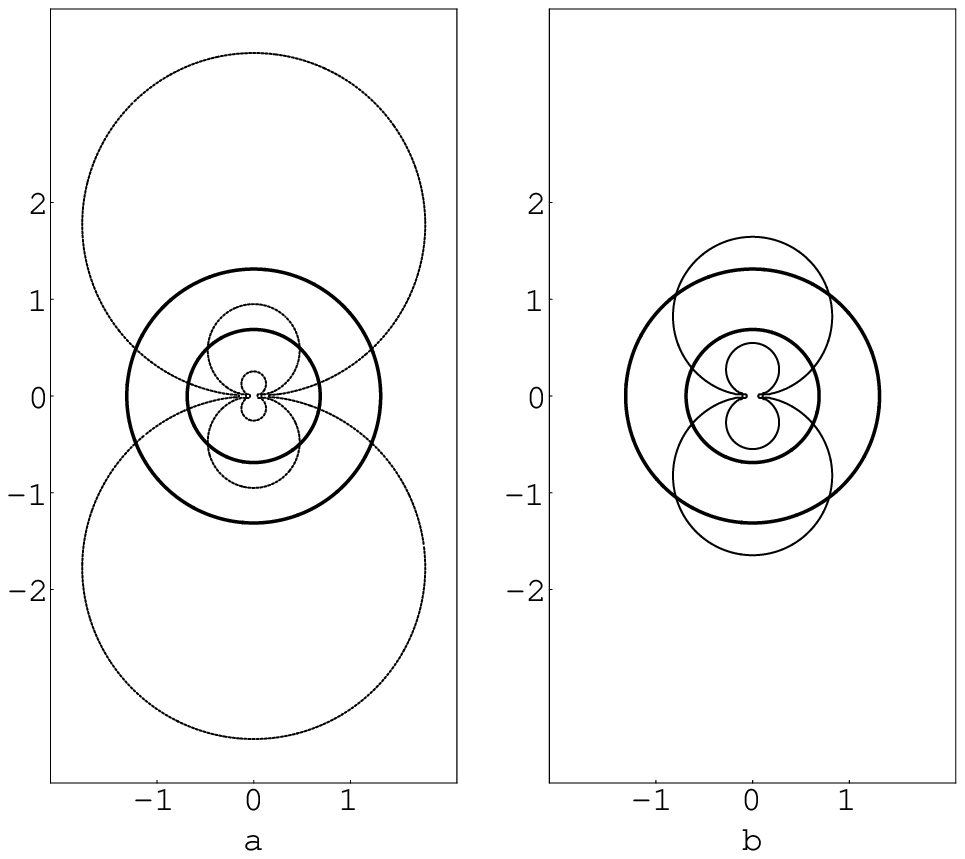}
\end{center}
\caption{
Kerr Black Hole with parameters $M=1$, $a=0.95$. Dashed lines in Fig. 3a correspond to both $K_1=0$ and $K_3=0$ ($K_1=-K_3$ in vacuum); thin solid lines in Fig. 3b correspond to $K_2=0$; thick solid lines in both Figs. 3a and 3b are the inner/outer horizons.
}
\label{fig3}
\end{figure}

\begin{figure}
\begin{center}
\epsfysize=5.0in \epsfbox{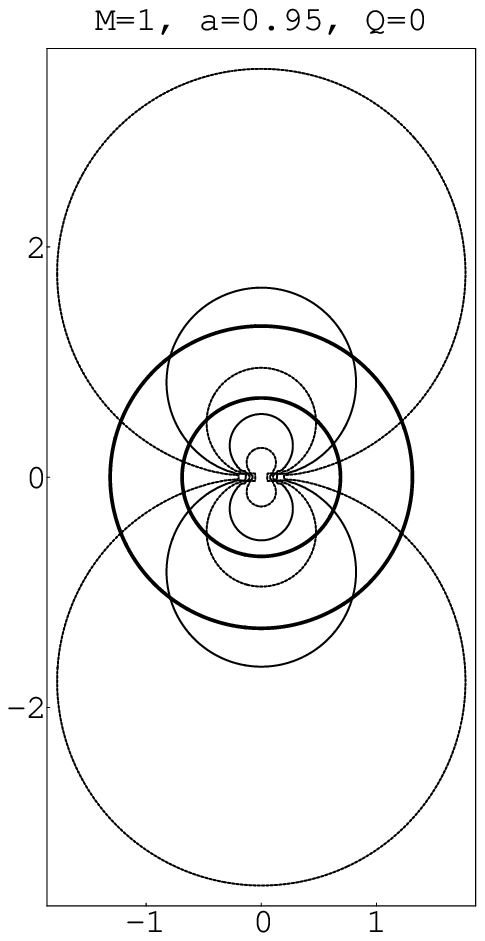}
\end{center}
\caption{
Kerr Black Hole with parameters $M=1$, $a=0.95$. Superposition of Figs. 3a and 3b. Dashed lines correspond to both $K_1=0$ and $K_3=0$ ($K_1=-K_3$ in vacuum); thin solid lines correspond to $K_2=0$; thick solid lines are the inner/outer horizons.
}
\label{fig4}
\end{figure}

\begin{figure}
\begin{center}
\epsfysize=2.8in \epsfbox{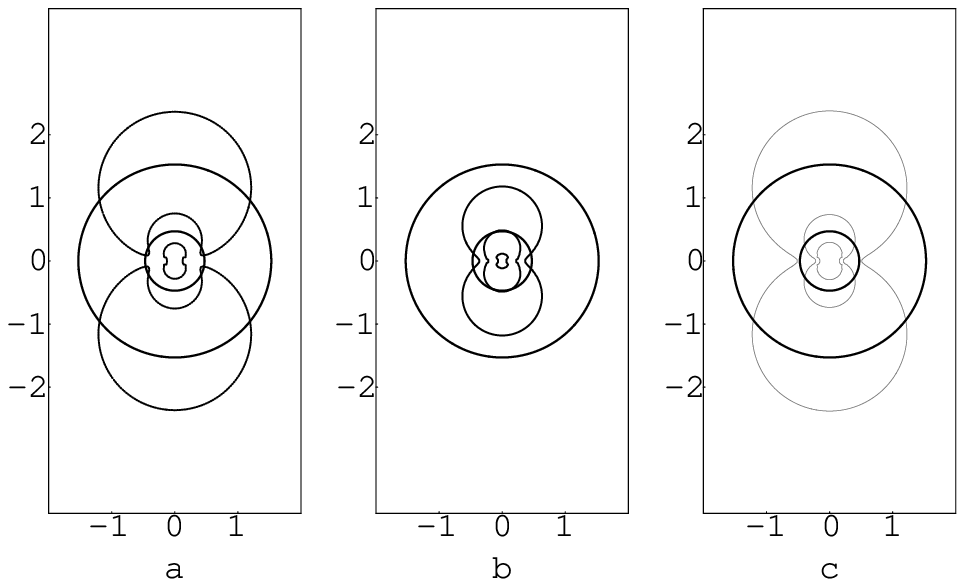}
\end{center}
\caption{
Kerr-Newman Black Hole with parameters $M=1$, $a=0.6$, $Q=0.6$. Dashed lines in Fig. 5a correspond to $K_1=0$, thin solid lines in Fig. 5b to $K_2=0$, grey solid lines in Fig. 5c
to $K_3=0$ and thick solid lines are the inner/outer horizons.
}
\label{fig5}
\end{figure}

\begin{figure}
\begin{center}
\epsfysize=5.0in \epsfbox{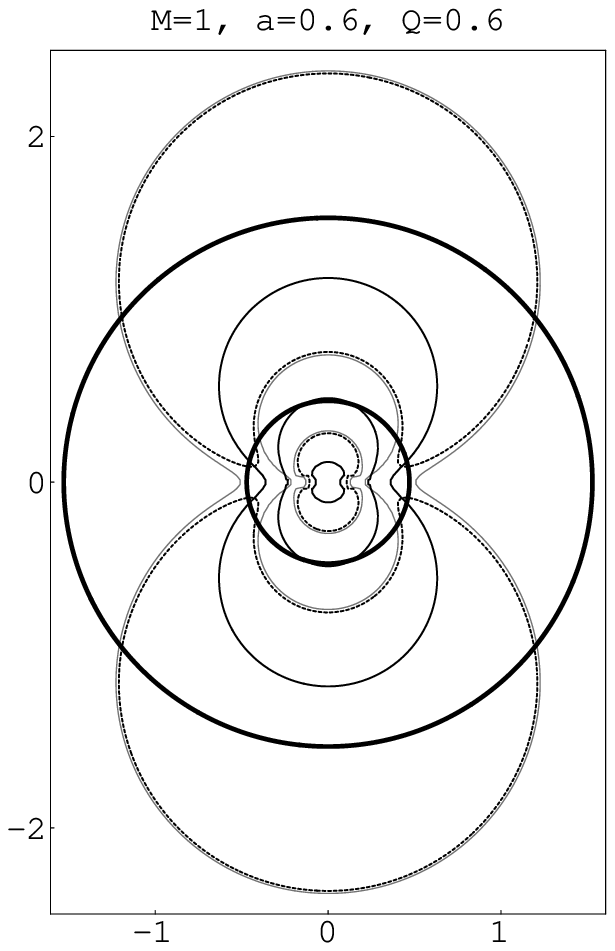}
\end{center}
\caption{
Kerr-Newman Black Hole with parameters $M=1$, $a=0.6$, $Q=0.6$. Superposition of Figs. 5a, 5b and 5c. Dashed lines correspond to $K_1=0$, thin solid lines to $K_2=0$, grey solid lines
to $K_3=0$ and thick solid lines are the inner/outer horizons.
}
\label{fig6}
\end{figure}

\begin{figure}
\begin{center}
\epsfysize=2.8in \epsfbox{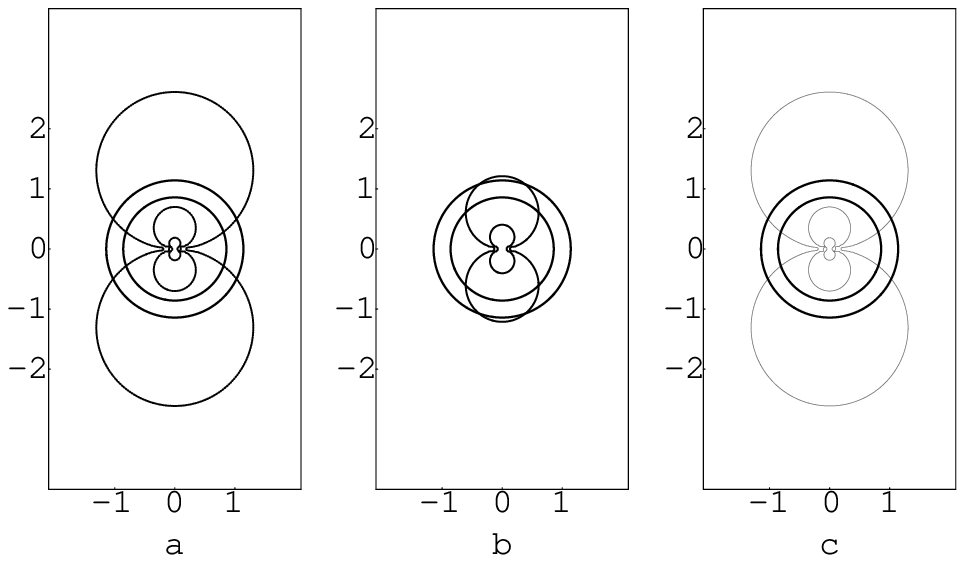}
\end{center}
\caption{
Kerr-Newman Black Hole with parameters $M=1$, $a=0.7$, $Q=0.7$. Dashed lines in Fig. 7a correspond  to $K_1=0$, thin solid lines in Fig. 7b to $K_2=0$, grey solid lines in Fig. 7c
to $K_3=0$ and thick solid lines are the inner/outer horizons.}
\label{fig7}
\end{figure}

\begin{figure}
\begin{center}
\epsfysize=5.0in \epsfbox{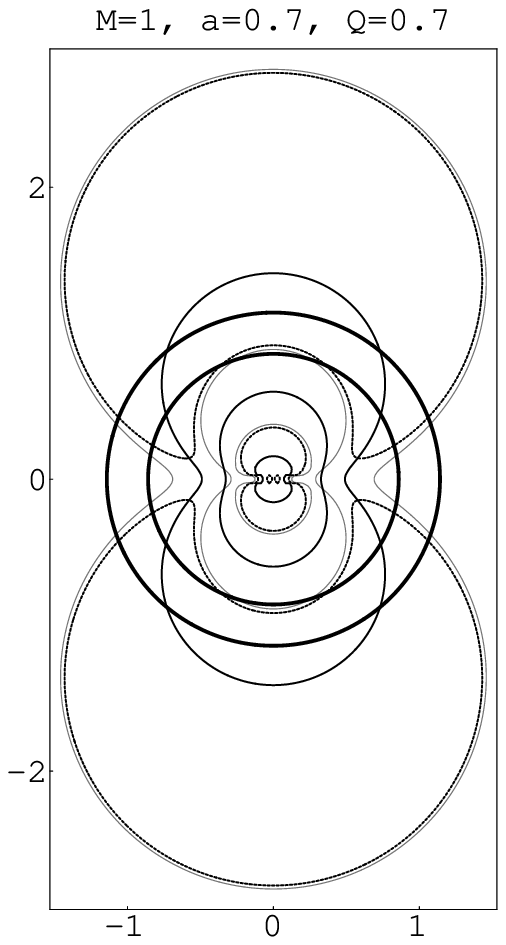}
\end{center}
\caption{
Kerr-Newman Black Hole with parameters $M=1$, $a=0.7$, $Q=0.7$. Superposition of Figs. 7a, 7b and 7c. Dashed lines correspond to $K_1=0$, thin solid lines to $K_2=0$, grey solid lines
to $K_3=0$ and thick solid lines are the inner/outer horizons.}
\label{fig8}
\end{figure}

Figs.~(\ref{fig1})--(\ref{fig4}) and (\ref{fig5})--(\ref{fig8}) show these surfaces  for the Kerr  and  Kerr-Newman cases respectively for two typical parameter choices. As usual in the literature the plots have been drawn in polar-like coordinates $(x,z)$, related to the Boyer-Lindquist coordinates $(r,\theta)$ by the relations
$x=r\sin\theta$ (horizontal axis), $z=r\cos\theta$ (vertical axis).
Note that in the $(x,z)$ coordinates, curves represented by the (polar) relation $r=\lambda \cos \theta $, (with $\lambda $ constant) as in Eqs.~(\ref{kerrzeros}) correspond to circles with center at $(0,\lambda/2)$ and radius $\lambda/2$.

We conclude this section by listing the gravitoelectromagnetic components of the Riemann tensor in the Kerr-Newman spacetime, which once inserted into eqs. (\ref{gravk}) allow an alternative (to NP) and straightforward derivation of the second order scalar invariants of the Riemann tensor.
The Carter observer family  with four-velocity
\begin{equation}
u_{\rm (car)}=\frac{(r^2+a^2)}{\sqrt{\Delta\Sigma}}(\partial_t+ \frac{a}{r^2+a^2}\partial_\phi)\, ,
\end{equation}
as expressed in Boyer-Lindquist coordinates,
where $\Delta=r^2+a^2-2Mr+Q^2$ and $\Sigma=r^2+a^2\cos^2\theta$, induces the $1+3$ decomposition of the orthonormal frame naturally associated with the Kinnersley null frame.\typeout{reference to Carter frame??}
One finds the following coordinate components with respect to the Boyer-Lindquist coordinates
\typeout{are these coordinate components or orthonormal frame components??}
\begin{eqnarray}
{\cal E}(u_{\rm (car)})_{tt}&=& \frac{a^2\sin^2 \theta}{\Sigma^4}[a^2(a^2\cos^2\theta)+Mr(r^2-3a^2\cos^2\theta)]\nonumber\\
&=& -\frac{a}{r^2+a^2}{\cal E}(u_{\rm (car)})_{t\phi}\nonumber \\
&=& \frac{a^2\sin^2\theta}{\Sigma^2}{\cal E}(u_{\rm (car)})_{\theta\theta}\nonumber \\
&=& \frac{a^2}{(r^2+a^2)^2}
{\cal E}(u_{\rm (car)})_{\phi\phi}, \nonumber\\
{\cal E}(u_{\rm (car)})_{rr} &=&-\frac{2Mr(r^2-3a^2\cos^2\theta)+Q^2(a^2\cos^\theta -3r^2)}{\Delta \Sigma^2}\, ,
\end{eqnarray}
and
\begin{eqnarray}
{\cal F}(u_{\rm (car)})_{\alpha\beta}&=&-{\cal E}(u_{\rm (car)})_{\alpha\beta}+\frac{2Q^2}{\Sigma \Delta}\delta^r_\alpha
\delta^r_\beta\, ,\nonumber \\
{\cal H}(u_{\rm (car)})_{\alpha\beta}&=&-\frac{a\cos\theta[M(a^2\cos^2\theta - 3r^2)+2rQ^2]}{Q^2(a^2\cos^2\theta-r^2)+Mr(r^2-3a^2\cos^2\theta)}[-{\cal E}(u_{\rm (car)})_{\alpha\beta}+\frac{Q^2}{\Sigma \Delta}\delta^r_\alpha\delta^r_\beta]\ .\nonumber
\end{eqnarray}

\section{Gravitoelectric or Gravitomagnetic Dominance in Vacuum Spacetimes}

For vacuum spacetimes the Riemann and the Weyl tensors coincide and the relation $K_1=-K_3$ holds.
As a consequence in a $1+3$ spacetime splitting point of view, the magnetic-magnetic part of the Riemann tensor
 reduces to its electric-electric part (apart from a sign, i.e. ${\cal F}(u)=-{\cal E}(u)$), while the electric-magnetic part ${\cal H}(u)$ becomes symmetric.
This is a simplified situation in which one has only two independent spatial tensor fields representing the Riemann tensor, exactly as in the electromagnetic case where the Maxwell 2-form $F$ is represented by the electric and magnetic vector fields; thus it is possible here to push forward the analogy between electromagnetism and gravitoelectromagnetism by considering the correspondence between the invariants of $F$ and those of $R$.

The detailed analysis of the Maxwell invariants in a curved background dates back to the seventies and to the pioneering
work of Ruffini, Hanni, Damour and Wilson.\cite{ruffini}
They introduced the concepts of a `region of electric dominance' (where for any observer $u$, $||E(u)||>||B(u)||$, i.e. $F_{\mu\nu}F^{\mu\nu}<0$), that of `region of magnetic dominance' (where for any observer $u$, $||E(u)||<||B(u)||$, i.e. $F_{\mu\nu}F^{\mu\nu}>0$) and that (stronger than the first two) of a `plasma horizon,' i.e. the boundary of a region in which the magnetic field can support an infinitely thin plasma against Coulomb attraction. In this case the meaning of the various regions in terms of trapping of particles by the magnetic field is clear.

When working out the analogy with the invariants of the Riemann tensor the most natural candidate to play the role of
$F_{\mu\nu}F^{\mu\nu}$ is $K_1$. One can rephrase the Ruffini et al definitions in terms of the corresponding gravitoelectromagnetic quantities
by introducing regions of `gravitoelectric dominance': $\Tr {\cal E}(u)^2>\Tr {\cal H}(u)^2 $, i.e. $K_1>0$, and regions of
`gravitomagnetic dominance': $\Tr {\cal E}(u)^2<\Tr {\cal H}(u)^2 $, i.e. $K_1<0$.
This gives a more reasonable interpretation of the regions where such a `quadratic' curvature becomes negative
rather that  claiming it to be `a new type of curvature' as does Henry.\cite{Henry}

In Figs. (1)-(4) the regions of gravitoelectric and gravitomagnetic dominance can be easily recognized. To this end one must consider the dashed circles which correspond to the vanishing of $K_1$: as a general feature, far from the hole the spacetime is gravitoelectrically dominated ($K_1>0$) while
a gravitomagnetically dominated region ($K_1<0$)  exists close to outer horizon (in other words, $K_1$ is positive at the spatial infinity  and changes sign at each crossing of the dashed circles).

Moreover, our discussion is in agreement with a pioneering work of de Felice\cite{fdf} which introduces the concept of `repulsive domains' in a curved spacetime (related to the definition of a `negative effective mass' for the black hole) as possible markers of the existence of a timelike singularity in terms of the vanishing of $K_1$ on surfaces containing the singularity itself.

Finally the analog of the concept of plasma horizon must be  substantially revised since Riemann tensor tidal forces act only on particles with intrinsic structure, like spinning test particles. In principle it is possible to attempt such a generalization but one must be careful in specifying what kind of particle/fluid trapping is under consideration.

\section{Conclusions}

The Kretschmann, Chern-Pontryagin and Euler invariants of the Riemann tensor have been written both in the NP and GEM formalisms for any spacetime in a form for which their evaluation can be done easily without the use of a computer algebra system, despite the widespread opinion in the literature to the contrary.
These scalars have been examined in the Kerr and Kerr-Newman cases.
The regions of spacetime where such objects vanish have been studied analytically, leading to the definitions of regions of gravitoelectric and gravitomagnetic dominance, at least for a vacuum spacetime. This approach is in complete analogy with the electromagnetic case and suggests an alternative interpretation of the scalar invariants.

\section*{Acknowledgements}
D.B. acknowledges warm hospitality at the "Istituto per le Applicazioni del Calcolo, M. Picone", CNR, Rome.

\newpage

\newpage

\newpage

\newpage

\newpage

\newpage

\newpage

\newpage

\end{document}